# Progress of concepts and processes in library information system: towards Library 2.0

Michael SCHERER (Conseil Général de la Moselle, France), Sahbi SIDHOM (LORIA/KIWI & Nancy Université, France)
e-Mails: michael.scherer@cg57.fr, sahbi.sidhom@loria.fr

*Abstract* — The main principle of the Library 2.0 is in the fact that the information has to be spread from the library to the user and vice-versa, to allow fast and permanent adaptation of the library services. Within the framework of the implementation of the "Departmental Plan of the Public Services Reading" by the "General Council of Moselle", the division of the public reading develops a departmental portal as main vector of the information with various users' profile.
The context of this research work takes a part of a Master degree training Diploma in STI-Economic Intelligence (Nancy2 University), combining facets of R&D in a professional context at the "Conseil Général de la Moselle" in France.

*Index Terms* — Library information system, economic intelligence, watch process, information needs, knowledge organisation.

## I. INTRODUCTION

Public needs change, manners change and the library has to answer to new practices. It is not any more question of giving physically documents to readers within library, media library or multimedia library infrastructures. We are crossed in the era of libraries 2.0 with the connection to Web 2.0. The major principle of the library 2.0, living in the fact that the information has to circulate from the library to the user and from the user to the library, is to allow a fast adaptation and perm services.

The last investigation on the cultural practices of French people confirmed the fact: "the number of subscribers in libraries decline; young people, turned to new numeric practices, they are less interest and don't come usually to library infrastructures". The library professionals wonder about the risk of obsolescence of the library equipments and think about the possible solutions.
- "*Should we speak about the concept of the digital library?*"
94 % of 18-24 yrs. people have a computer [1].
- "*Have they still reasons for frequenting a library?*"
At the same time, the public reading is in the target of most current stakes in the public politics.
It is enough to see actually numbers of projects to create new libraries or media libraries with the fact "No interest" shown by cities for these infrastructures: It is the paradox about the actual situation.
User's notion is consequently very important in the implementation of a multimedia library system. Users must be considered as consultants, participants, co-creators; this is to allow the innovation as well in the virtual services as the physical services of a library or a media system.

Within the framework of the implementation of the "Departmental Plan of Public Reading" [2] by the "General Council of the Moselle", the Service of the Development of Territories and the Public (SDTP), under Pascale Valentin-Bemmert's responsibility, as Project Manager, develops a departmental portal of the public reading. In this context, the main reflections turn around the following axes, as a set of questions "*How to...?*":
– Go towards the user;
– Open to the user, that he becomes "a contributor";
– Improve the existing tools with the deepening of the traditional missions;
– Propose new services with the diversification of the traditional missions.

These questions are proposed in the objective of this library 2.0 portal. It is therefore to understand the stakes of this portal, to know the decision maker and his objectives, to understand the decision problem and to convert it into information retrieval problem. Also, to think about the information integration in the library 2.0 system, their updates and about the collaborative aspects to define: "*What will have access to the user?*".

It is this report problem that concerns the problematic of our paper. This paper is articulated around three research dimensions, namely: the User, the Information and the Knowledge. We integrate a modelling work on the library information system opened to the multimedia objects.

In a first part, section II, we present the methodology and the implemented tools. Then in section III, we present the results obtained with their impacts. At last section, we approach the perspectives of the library 2.0 information systems.

We want our readers to observe that this work has been done in the context of a Master degree training Diploma in STI-Economic Intelligence (Nancy2 University), combining facets of R&D in a professional context at the "Conseil Général de la Moselle" (Metz) in France.

## II. METHODOLOGY AND TOOLS

Since the Internet networks, the quantity of information doesn't stop increasing and it's important to target well these information resources to propose the most relevant. It is a question to:

*i)*- Qualify the resources to organize routes (courses) in the mass of the contents:
- Make readable the contents,
- Help to find for what we do not look: leave information, re-cut it, develop it, etc.
- Show the professional competences and recommendations.

*ii)*- Build with the user:
- Comments: suggestions, requests and preferences of users,
- New information contributions.

To anticipate at the best our reflections (*i, ii*), we set up a methodology applied to the study dimensions, as well as the working tools.

### A. Applied Methodology

**"How to organize the courses of the users in the mass of the library resources?"**

The complexity of this problem lives in the fact that several axes of reflections are to be taken into account. We shall speak about three dimensions for the construction of a coherent information system [3] (**Fig. 1**):
a) the dimension of "actors" (or users: U);
b) the dimension of "information" (I) which circulate and to capitalize;
c) the dimension of "knowledge" (K).

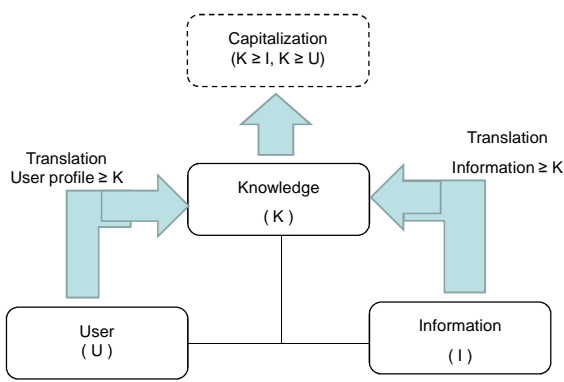

**Fig. 1:** Representation of dimensions: User, Information and Knowledge [4].

An association takes place in these three dimensions (cf. **Fig.1**), namely:
- Documents and their diversity: from text to multimedia;
- Processes and interoperability in the indexing and in knowledge management;
- Users and their information needs: the indexer, the watcher and the decision maker through the sharing processes (information, knowledge and collaboration).

We can separate this representation in two parts:
- Construction/Modeling with links and transitions between actors, information and knowledge;
- Exploitation, visible part by the public (H-M interface).

The information integration by the user and the transition of information to knowledge is an important phase in a (multimedia) library information system. This phase is made with the importance given to the transitions with the "kernel" about actors, information and knowledge. In this project, the notions of information validation by the user, the accessibility to information are also taken into account.

The document indexing process is an important stage in the implementation of a library system and the information retrieval. If the indexation is not relevant, the document will not be found with regard to a user query. It is important to transform and to find communication symmetry between the natural language and the documentary language. Today, it is useful to take into account the index descriptors in the form of (social) tags or a set of referenced words (in thesaurus), which are realized by users directly and with their own terminology.

In the information dimension, the contents and the capacity to find it in a most coherent way are essential keys of the system.

### B. Working Tools

Within the framework of the implementation of a multimedia library information system, we evolve on three dimensions: user (U), information (I) and knowledge (K). Each dimension has its importance because of its implication in the functional architecture of the library system.

<u>User Dimension</u>

The exploration of the user axis, we proceed to a modelling of the decision-maker and its environment. It is to make explicit the knowledge of the decision-maker and about the knowledge of his environment. We are going to meet on a structure a set of knowledge about the decision-maker and knowledge about the organization environment (library infrastructure).

The decision-maker in the internal and external environment of the organization: "*who is capable of identifying and raising the problem to be resolved into term of stake, risk or threat which presses on the company*". In other words, he knows the needs of his company, the stakes and possibly the risks.

In our reflection, the environment of the decision-maker is rather particular because it gets organized around a complex organization of the "Division of the Public Reading" and Libraries. This division consists of 4 poles distributed on the "Mosellan territory" all around of which weaves 135 libraries.

The Department of the "Moselle" chose, via the Departmental Plan of Reading on Public 2009-2014 to decentralize all the missions and the services of public reading. The notion of portal as collaborative space can put in the Direction of the Resources and the Computing Services of the problems of access and security. It is one of notions to be taken into account at the time of the study of the proposed library information systems by companies. Indeed, the current IT environment is complex with centralized IT architecture and security constraints which, at the moment do not allow us to reach in certain "collaborative" sites (like facebook) and sites of audio, videos… resources.

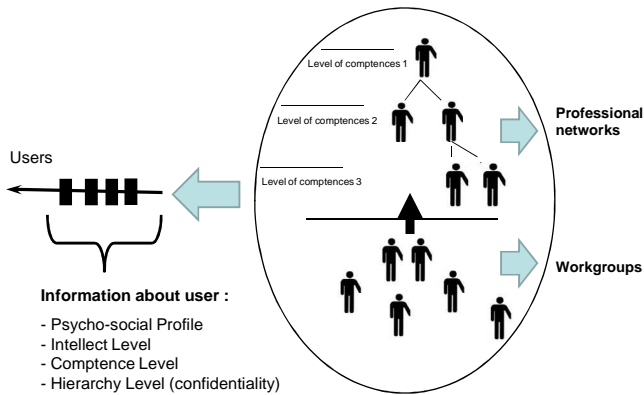

**Fig. 2:** User dimension: actor in the organization.

In the library environment, the decision-maker knowledge allows us to know other elements to implement the portal, such as the questions of homogeneity in the communication, the respect for the graphics standards and the integration of the portal in the public reading within the portal of the CG57.

In this information search process, it is fundamental to establish a profile of the user, to treat the information and establish the representation of documents to users (**Fig. 2**).

There is a big variety of users. Every user is going to express different and very specific needs. It is important to know the level of knowledge and the preferences of user. The needs of the users are often expressed in natural language. The difficulty lives in its transformation (ie. processing) in request. In this case, we can observe the recommendation models and collaborative filtering to get the profile and user preferences in the system.

The users' information will feed the user database.

To acquire this information, we can use three methods:
- Explicit model: it consists in asking explicitly the user for his preferences. It is the simplest method;
- Implicit model: it consists in observing the user activities in search situation (to find his information tracks);
- Mixed model: it consists to combine static and dynamic methods about user activities and profile.

In library infrastructures with feasibility reasons, we shall privilege the explicit model for the information user acquisition as well as library professionals are listening to their public.

Information dimension

In the flow of continuous information, it is necessary to establish a methodology of information search, by taking into account the stratification of the information: primary resources (documents), the secondary resources (bibliographic records) and tertiary resources (user annotations, social tags...).

In a consideration of contents supply, we can ask the question about: "*what information we want to make available to the user?*", to determine the types and information sources collected or made.

Within the framework of multimedia information system such as in the library portal, the information types should be diverse: text, photos, videos, (musical) audio extracts... and resources according to their relevance in Google books, wikipedia, lastfm, BnF, Gallica, etc.

In primary information, we can consider resources in the open databases, open archives, Internet. At this level, tools proposed by other poles of public reading reflect the features about the use, the relevance of the services and tools' ergonomy proposed to the user. With the progress of the information contents of towards the Web 2.0 tools (like as multimedia tools): blogs, flow RSS, podcast, webradios, netvibes, digitalizations (Gallica resources). These sources allow to complete the information according to the user needs (**Fig. 3**).

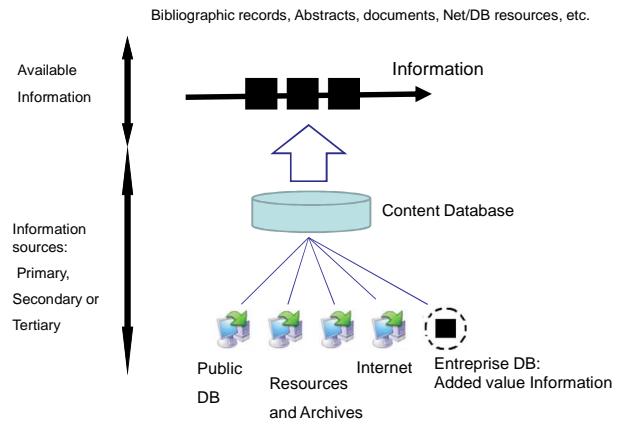

**Fig. 3:** Information dimension: information resources in the organization.

Knowledge dimension

To be able to establish footbridges and comparative degrees between the existing information and the one that we wish to propose, we set up a watch process. The watch process consists in giving the pertinent information to the target user, at the right time, to make the good decision.

"*What information? to whom? and how?*". From these questions, we can establish a projection between the exploitable information and the nature of the knowledge proposed to the user. This projection is based on the treatment and the analysis of the information through the watch process.

In front of the increasing mass of information, the libraries need systems of assistance to analyse more and more successful. These tools have to offer possibilities of very fine exploration and synthetic representation of the meditative information and the new generated knowledge observed in the processing. So, watch tools did not stop perfecting to allow a follow-up reasoned by this strategic information flow, by automating partially the process. The contribution of the watch actors will also be integrated into this process.

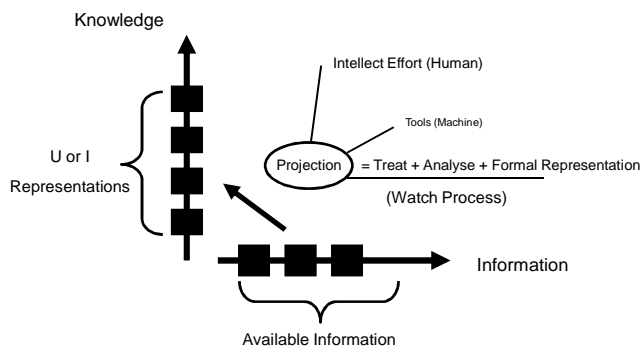

**Fig. 4:** Knowledge dimension: conceptual entities in processing.

The information corresponds to the human interpretation on the raw data. The information arises from resources grouping. About primary information, it produced by the authors, secondary information (or bibliographical notes) produced by the information officers and the professionals of libraries, and tertiary information (annotations, tags and votes of users) produced by every user who lands in system interface (**Fig. 4**).
The knowledge is information crossings by process, interpretation by user in the information models which give meaning and makes it operational (meaningful in the system).

In the system implementation context, we attempted to define how user can access to product resources, how this information is transformed into knowledge, how user access to knowledge and how he enriches the contents: it is about the re-indexing contents by user activities [5].
The user became an author. He can, by acquainting with the document, bring it additional and complementary information. As annotator, this new actor develops informative contents by integrating "new information, links, and interpretative elements on the document" [6].
The annotation indicates at the same time the activity of the additional new information on a document and the result of the annotator action. In a common way, a note (by annotation) is a brief comment, an explanation, etc. about the document or about its content [7].
From a general point of view, notes are well appreciated by readers. A comment, an appreciation or a critic is going to call to the future reader, who via these notes is going to find an interest for the annotated document, is going to want to build up to himself its own opinion (notice).

This library 2.0 portal is in progress in a collaborative way between the agents of the DLPB, the mosellans librarians, the users via forums or comment zones, and must be daily updated. These new "places" of exchange and sharing have for purpose: "to support cooperation activities in more structured way, in which the interactions also lean on information or documents, shared by a collective pursuing with common objectives" [8].
It exists various relations between the transmitter (Library professionals) and the receiver (Library users):
- Common project in the main objectives or in the partially shared interests;
- Social relationships between the professional actor and the beneficiary user to ensue partly the interests or the common objectives.

However, filtering techniques on information must be organized to verify the coherence and the relevance of certain information additions.
The enrichment of information databases by users brings an added value to the information system as far as this one is relevant. The library 2.0 information system has to take into account the information about the environment in which it is going to be used and which will be in its objectives.

In next, the objective is going to involve in the model proposed to the decision-making problem and so to our problem in the Economic Intelligence (EI) process.

## III. RESULTS

More than a simple Website of contents, the library 2.0 portal includes at least application resources and complementary services. It has a double mission of resource aggregator and dynamic menu.
As aggregator of resources, it combines in the same frame of the information stemming from diverse and heterogeneous systems. As dynamic menu, it presents a state of the available resources and offer of the links towards information.
Others proposed services, the appropriate added value in the portal is the selection and the meeting, from a certain number of tools and resources: contents editorial, search engine, diverse information, news, classification of Websites by themes, e-mail, forums, etc.
Also, the portal integrates a dimension of customization and identification of the Internet users according to their various objectives. So, it allows the Internet users to define a personal space, to place selections, requests, alerts which they wish to define, etc.

The objective of this departmental portal of reading is to propose a big panel of services from a single address. The portal has to enter the participative Web (or Web 2.0) and so turn to the major principles of the library 2.0. With the real strategy of services 2.0, it is important to locate as development vector of the new technologies within libraries, and to associate it all the participants of the departmental network. The librarians and the Internet users, in Web 2.0 with IT infrastructures, can add opinions (notices) on a book, a movie record next, comments or to import extern source on it.

For resources, every library 2.0 user can index by proposing new tags or keywords besides those existing. It is a question of enriching the possibilities about discovering books, records and videos... [9]. The notes on books, records and videos will integrate the library information system.

Finally, the portal allows an access to the common "catalogue" (library user interface) of all libraries in Internet network: in autonomous or guided way, user navigates via tags, themes' lists, books' categories, videos' collections… (**Fig. 5**).

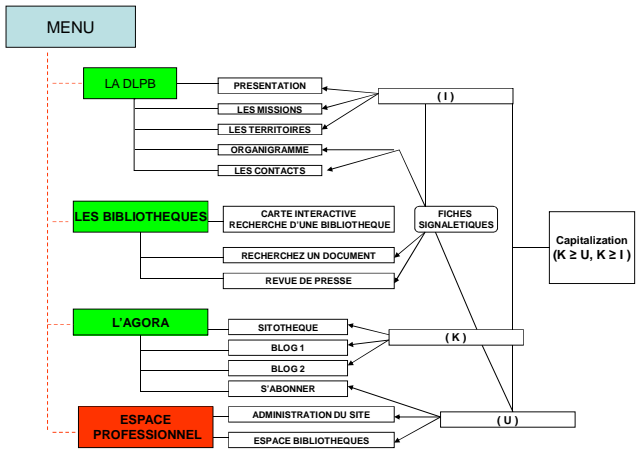

**Fig. 5:** "Departmental portal of public reading" architecture.

Application dimension

The departmental public portal of reading of the Moselle should be born during the 1st half-year 2011. This very ambitious project was launched more than a year ago through the constitution of a workgroup which has for mission to work on a reflection relative to the use of this portal (**Fig. 6**).

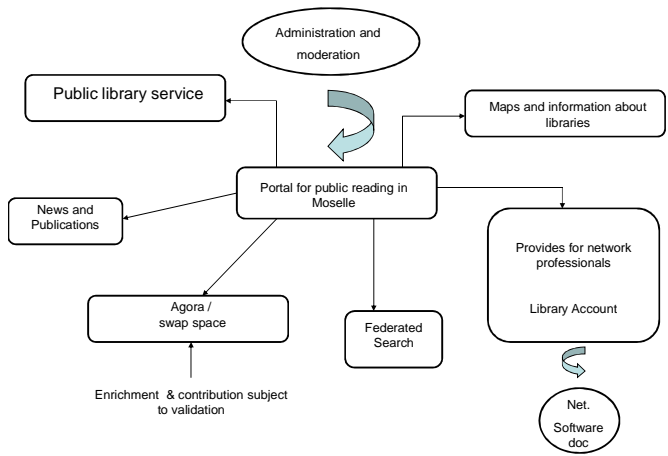

**Fig. 6:** Application plan of the departmental portal of reading.

Several points will already be to respect, at the level of the Graphics standards in particular. It is important to have viewed well the objectives, the missions and the constraints of the portal to estimate the work to be led for its realization and for its functioning (update, maintenance).

It will be a question of proposing the most relevant information by the actions of users as the library actors. The user is in the target of this portal. The portal is a shop window of the politics of the Department in public reading. It will be the interface between the professionals of the reading, the libraries and the Internet users.

IV.  CONCLUSION AND PROSPECT

Today, the roles were inverted. The reader does not come to look any more for the information, it is the information which has to come to him. The library infrastructure is not any more the place of information access. The user does not want to have any more customer's simple role, passive way. He wants to be a contributor, an actor, to express his opinion, to participate actively in the library life.

Also, today, the information is exchanged and shared by Internet networks. The library, via its Website or its portal has to allow the user to reach its resources from his home (or connected WIFI place).

The interaction with the library and user needs is in the target of this library information systems. The library has to renew its services, see again (and revise) its missions. Without this objective, the Division of the Public Reading and the Library politics could see it in restriction. The proposed departmental portal is going to bring by the 1st half-year 2011 these new services: this new vision of librarian's job which turns to the tools of Web 2.0 to attract new public, to develop new adapted services (**Fig. 7**).

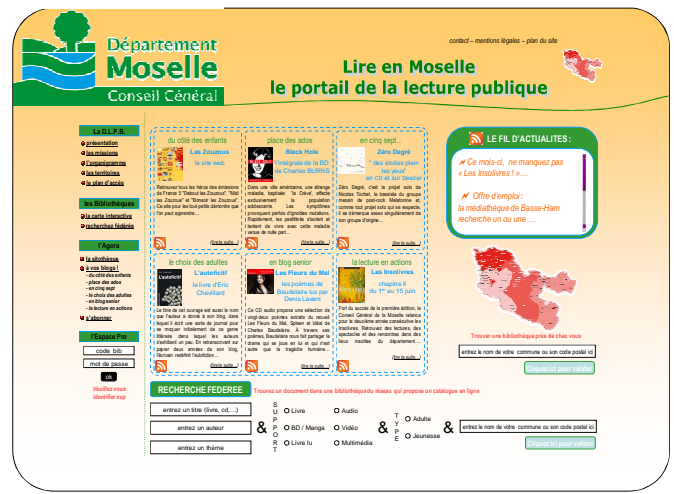

**Fig. 7:** The new departmental portal of reading.

It is necessary in a short time to progress from BDP (in French: "Bibliothèque Départementale de Prêt") as "traditional" type of library, whitch its main function was the contribution on collections and the document services, to a BDP as "modern" type, which the digital technology and where the e-resources (multimedia documents) contribution are the main activity with user and professional.

## V. BIBLIOGRAPHY

[1] R. Bigot, P. Croutte. (2009). La diffusion des technologies de l'information et de la communication dans la société française. in Rapport réalisé à la demande du Ministère de l'Economie, de l'Industrie et de l'Emploi, novembre 2009.

[2] P. Valentin-Bemmert. (2008). Schéma Départemental des Services de Lecture Publique 2009-2014. in Document de travail présenté lors de la Commission permanente du Conseil Général de Moselle. décembre 2009.

[3] S. Sidhom. (2010). Conception de systèmes d'information multimédia. in Cours en Master 2 IST-IE, Université de Nancy2. Oct. 2010.

[4] S. Sidhom. M. Ghénima. P. Lambert. (2010). Systèmes d'information et Intelligence économique : enjeux et perspectives. in Proceedings Colloque international IEMA-4, Alger, mai 2010. (S. Sidhom guest speaker).

[5] A. Harbaoui, M. Ghenima, S. Sidhom. (2009). Enrichissement des contenus par la réindexation des usagers: un état de l'art sur la problématique. in Proceedings Conférence Internationale SIIE 2009. vol.1(2009). pp.932-942, IHE éditions Tunis.

[6] S. Sidhom. (2008). Approche conceptuelle par un processus d'annotation pour la représentation et la valorisation de contenus informationnels en Intelligence économique (IE). in Proceedings Conférence internationale SIIE'2008. Vol.1, pp.172-190, IHE éditions Tunis.

[7] P. Kislin. (2007). Modélisation du problème informationnel du veilleur dans la démarche d'Intelligence Economique. Thèse de doctorat en Sciences de l'Information et de la Communication, Université Nancy 2.

[8] M. Zacklad. (2005). Processus de documentation dans les Documents pour l'Action (DopA) : statut des annotations et technologies de la coopération associées ». in Proceedings colloque « Le numérique : Impact sur le cycle de vie du document pour une analyse interdisciplinaire », 13-15 Octobre 2004, Editions de l'ENSSIB, Montréal (Québec).

[9] D. Desmottes. (2010). S'approprier les outils de l'information et de la communication. CNFPT-Technologies et bibliothèques, ENACT de Nancy, septembre 2010.

[10] O. Ertzscheid. (2006). Etude exploratoire des pratiques d'indexation sociale comme une renégociation des espaces documentaires. Vers un nouveau big-bang documentaire ? in ADBS Éditions, *2006*. 344p. Collection Sciences et techniques de l'information. Fribourg: (*2006*)

[11] Chartron G. (2008). « L'offre documentaire numérique : repères et décryptages », Documentaliste – Sciences de l'information, n°2, mai 2008.

[12] P. Lambert (2009). ChroniSanté : Un système d'information d'aide à la décision pour la prise en charge des patients atteints de maladie chronique. Mémoire de Master IST-IE, Université Nancy 2. 2009.

[13] P. Lambert, S. Sidhom (2010). Extraction des connaissances et visualisation: cas d'application sur un corpus du projet *ChroniSanté* en France. in Proceedings Conférence Internationale SIIE 2010. vol.1(2010). IHE éditions Tunis.

[14] S. Sidhom (2002). Plate-forme d'analyse morpho-syntaxique pour l'indexation automatique et la recherche d'information : de l'écrit vers la gestion des connaissances. Thèse de doctorat, université Claude Bernard, Lyon 1.

[15] P. Félizart-Chartier. (2007). La bibliothèque fait de société, mutiservices, nouveaux services. Atelier « l'avenir des bibliothèques, les bibliothèques de l'avenir », journées nationales de l'ADBDP, 2006.